\documentclass[12pt]{iopart}

\usepackage{bm}
\begin{document}

\title[]{On the origin of preferred-basis and evolution pattern of wave function}

\author{Shizhong~Mei}

\address{Micron Technology, Boise, Idaho 83707, USA}
\ead{smei@micron.com}
\begin{abstract}
The standard quantum mechanics assumes Schr\"odinger equation for regular evolution and wave function collapse for measurement. As shown in this paper, only particular collapse equation can continuously transition to Schr\"odinge equation. The collapse equation also adds some restriction to the preferred-basis. Under the assumptions that the preferred-basis depends on the whole system Hamiltonian but is not affected by the weights of the basis functions in the system wave function, a unique set of determination equations of the basis functions is derived from the collapse equation. The second order time derivative of the wave function is continuous at the end of the collapse. To make the derivative continuous at the beginning of the collapse, it is proved that the collapse equation has to contain a cyclic function with period twice the duration of the collapse, which leads to conditioned alternating Schr\"odinger evolution and collapse of equal duration.

\end{abstract}

\pacs{03.65.Ta, 03.65.-w}
\vspace{2pc}
%\noindent{\it Keywords}: Wave function collapse, Schr\"odinger equation, preferred-basis, Generalized eigen-equations, Duration of collapse
\maketitle

\section{Introduction}
According to the standard quantum mechanics [1], the evolution of a wave function is subject to dichotomous laws---the wave function undergoes regular evolution described by Schr\"odinger equation before and after a measurement; during the measurement, however, the governing equation changes to other types; the result of the second type evolution is probabilistic and in any event the end state is an eigenfunction of the observable being measured. This theory has achieved great success in predicting experimental results. Nevertheless, the unknown mechanism to determine preferred-basis, the unknown condition for a collapse to occur, and the missing dynamics of collapses have motivated research now and then [2]. 

One category of the research, e.g., [3-10] focused on the interaction between a sub-system and its environment and the work has led to a decoherence program. In that program the authors demonstrated that the interaction could result in near diagonalization of reduced density matrices associated with a few preferred-basis. It also offered an explanation on the origin of those bases by using particular decomposition of the universe into object being measured, measuring apparatus, and environment. Because the decomposition into apparatus and environment is unjustified, the explanation is problematic. Additionally, that program does not provide the expression for preferred-basis functions that are, e.g., quasi-position eigenfunctions. Besides, the accuracy of the decoherence condition needs to be improved. In another category of the research such as [11-18], the authors added a few terms to Schr\"odinger equation to reduce the wave function stochastically. Although those models are intuitively appealing, they do not generally achieve exact end states with exact probabilities.

So far there has not been any report on analyzing the transition between Schr\"odinger equation and the missing collapse equation in the standard quantum mechanics. As will be shown in this paper, for the dynamic equations to continuously transition from one kind to the other, the collapse equation needs to take specific form that in turn puts some restriction on the preferred-basis. The second order time derivative of the wave function is continuous at the end of the collapse. In case the second order time derivative of the wave function is also continuous at the beginning of the collapse, analysis in this paper shows that the collapse equation has to contain a cyclic function with period twice the duration of the collapse. The periodic function suggests that the two dynamic equations take turn to guide the wave function of the universe.

The restriction from the collapse equation, together with the assumptions that preferred-basis depends on the whole system Hamiltonian but not on the weights of the basis functions in the system wave function, leads to a unique set of equations that determines the preferred-basis for general situations. A few asymptotic solutions solved from the determination equations include approximate energy eigenfunctions, quasi-position eigenfunctions, mixed energy eigenfunctions and quasi-position eigenfunctions. Those derived preferred-basis provide satisfactory explanations on the results of many quantum measurements. 

The purpose of this paper is to provide details on the analysis and derivation of the missing collapse equation and the unique determination equations of preferred-basis. The rest of the paper is organized as follows. Section 2 analyzes the properties of the missing collapse equation. Section 3 analyzes the restriction of the collapse equation on preferred-basis. Section 4 derives the unique set of equations that determines preferred-basis for general systems. Section 5 proves the periodicity of the functions contained in the collapse equation. Discussions and conclusions are provided in Section 6 and 7 respectively.

\section{The missing collapse equation}
\subsection{The basics of the collapse equation}

The analysis of this paper is confined in the framework of the standard quantum mechanics. That's to say, for a system with continuous wave function $\mid\psi(\mathbf{x},t)>$ and Hamiltonian $H(\mathbf{x},t)$, the wave function $\mid\psi(\mathbf{x},t)>$ satisfies Schr\"odinger equation
\begin{equation}
\setlength\abovedisplayskip{0pt}
\setlength\belowdisplayskip{0pt}
i\hbar\frac{\partial}{\partial{t}}\mid{\psi}(\mathbf{x},t)>=H(\mathbf{x},t)\mid\psi(\mathbf{x},t)>
\end{equation} 
before and after a collapse, which refers to some process that finishes with the state of a sub-system being one of the preferred-basis functions. Since generally Schr\"odinger equation does not collapse the wave function, the dynamics of a collapse will take a different form
\begin{equation}
\setlength\abovedisplayskip{0pt}
\setlength\belowdisplayskip{0pt}
i\hbar\frac{\partial}{\partial{t}}\mid{\psi}(\mathbf{x},t)>=\mid\phi(\mathbf{x},t)>+H(\mathbf{x},t)\mid\psi(\mathbf{x},t)>
\end{equation} 
where $\mid\phi(\mathbf{x},t)>$ is to be determined. It is worth noting that because of the arbitrary term $\mid\phi(\mathbf{x},t)>$, (2) represents a general equation. 

Assume that the preferred-basis functions $\mid\psi_{1,k}(\mathbf{x}_1)>$ for $k=1,2,\cdots$ are selected for sub-system 1 and that the collapse equation (2) holds for interval [$\tau$ $\tau+T$]. In order to avoid abrupt change of the wave function $\mid\psi(\mathbf{x},t)>$, the duration $T$ has to be non-zero. Due to the completeness of $\mid\psi_{1,k}(\mathbf{x}_1)>$ in the $\mathbf{x}_1$ domain, $\mid\phi(\mathbf{x},t)>$ can be expressed as
\begin{equation}
\setlength\abovedisplayskip{0pt}
\setlength\belowdisplayskip{0pt}
\mid\phi(\mathbf{x},t)>=\sum_k\mid\psi_{1,k}(\mathbf{x}_1)><\psi_{1,k}(\mathbf{x}_1)\mid\phi(\mathbf{x},t)>
\end{equation} 
where $<\psi_{1,k}(\mathbf{x}_1)\mid\phi(\mathbf{x},t)>=\int\psi^*_{1,k}(\mathbf{x}_1)\phi(\mathbf{x},t)d\mathbf{x}_1$. 

The inner products $<\psi_{1,k}(\mathbf{x}_1)\mid\phi(\mathbf{x},t)>$ are functions of $\mathbf{x}_{1\textrm{-}}$ or the coordinates of the objects outside the sub-system 1. For convenience, that group of objects is named sub-system 1-. Introducing functions $\lambda_k^{-1}(\mathbf{x}_{1\textrm{-}},t)$ such that 
\begin{equation}
\setlength\abovedisplayskip{0pt}
\setlength\belowdisplayskip{0pt}
<\psi_{1,k}(\mathbf{x}_1)\mid\phi(\mathbf{x},t)>=i\hbar\lambda_k^{-1}(\mathbf{x}_{1\textrm{-}},t)<\psi_{1,k}(\mathbf{x}_1)\mid\psi(\mathbf{x},t)>,
\end{equation} 
the collapse equation (2) becomes
{\setlength\arraycolsep{0pt}
\begin{eqnarray}
i\hbar\frac{\partial}{\partial{t}}\mid{\psi}(\mathbf{x},t)> =&i\hbar\sum_k\lambda_k^{-1}(\mathbf{x}_{1\textrm{-}},t)\mid\psi_{1,k}(\mathbf{x}_1)><\psi_{1,k}(\mathbf{x}_1)\mid\psi(\mathbf{x},t)>\nonumber\\
&+H(\mathbf{x},t)\mid\psi(\mathbf{x},t)>. 
\end{eqnarray}

Again (5) is a general equation. Meanwhile since (5) represents a physics equation, similar to the requirement on other established physics equations, functions $\lambda_k^{-1}(\mathbf{x}_{1\textrm{-}},t)\mid\psi_{1,k}(\mathbf{x}_1)><\psi_{1,k}(\mathbf{x}_1)\mid\psi(\mathbf{x},t)>$ and $H(\mathbf{x},t)\mid\psi(\mathbf{x},t)>$ need to be continuous and finite. 

According to the definition of the collapse, the end state contains only one basis function e.g., $\mid\psi_{1,\tilde{k}}(\mathbf{x}_1)>$, i.e.,
\begin{equation}
\setlength\abovedisplayskip{0pt}
\setlength\belowdisplayskip{0pt}
\mid\psi(\mathbf{x},\tau+T)>=\mid\psi_{1,\tilde{k}}(\mathbf{x}_1)><\psi_{1,\tilde{k}}(\mathbf{x}_1)\mid\psi(\mathbf{x},\tau+T)>.
\end{equation} 
At the beginning and the end of the collapse, the dynamics is both Schr\"odinger equation and the collapse equation, meaning 
\begin{equation}
\setlength\abovedisplayskip{0pt}
\setlength\belowdisplayskip{0pt}
\lambda_k^{-1}(\mathbf{x}_{1\textrm{-}},\tau)=0
\end{equation} 
for general situation where $<\psi_{1,k}(\mathbf{x}_1)\mid\psi(\mathbf{x},\tau)>\neq 0$ and
\begin{equation}
\setlength\abovedisplayskip{0pt}
\setlength\belowdisplayskip{0pt}
\lambda_k^{-1}(\mathbf{x}_{1\textrm{-}},\tau+T)\mid\psi_{1,k}(\mathbf{x}_1)><\psi_{1,k}(\mathbf{x}_1)\mid\psi(\mathbf{x},\tau+T)>=0.
\end{equation} 

\subsection{The resultant collapse equation}

For $k\ne\tilde{k}$, multiplying $<\psi_{1,k}(\mathbf{x}_1)\mid$ to both sides of (5) yields the following inhomogeneous equation 
{\setlength\arraycolsep{0pt}
\begin{eqnarray}
i\hbar\frac{\partial}{\partial{t}}<\psi_{1,k}(\mathbf{x}_1)\mid{\psi}(\mathbf{x},t)>=&i\hbar\lambda_k^{-1}(\mathbf{x}_{1\textrm{-}},t)<\psi_{1,k}(\mathbf{x}_1)\mid\psi(\mathbf{x},t)>\nonumber\\
&+<\psi_{1,k}(\mathbf{x}_1)\mid{H}(\mathbf{x},t)\mid\psi(\mathbf{x},t)>.
\end{eqnarray}
The solution is
\setlength\abovedisplayskip{0pt}
\setlength\belowdisplayskip{0pt}
\begin{eqnarray}
\fl <\psi_{1,k}(\mathbf{x}_1)\mid\psi(\mathbf{x},t)>=&e^{\int_\tau^t\lambda_k^{-1}(\mathbf{x}_{1\textrm{-}},t)dt}(
<\psi_{1,k}(\mathbf{x}_1)\mid\psi(\mathbf{x},\tau)>+\nonumber\\
&\int_\tau^t(i\hbar)^{-1}e^{-\int_\tau^t\lambda_k^{-1}(\mathbf{x}_{1\textrm{-}},t)dt}<\psi_{1,k}(\mathbf{x}_1)\mid{H}(\mathbf{x},t)\mid\psi(\mathbf{x},t)>dt).
\end{eqnarray}
Because of (6), either 
\setlength\abovedisplayskip{0pt}
\setlength\belowdisplayskip{0pt}
\begin{eqnarray}
<\psi_{1,k\neq\tilde{k}}(\mathbf{x}_1)\mid\psi(\mathbf{x},\tau)>+\int_\tau^{\tau+T}(i\hbar)^{-1}e^{-\int_\tau^t\lambda_{k\neq\tilde{k}}^{-1}(\mathbf{x}_{1\textrm{-}},t)dt}\nonumber\\
<\psi_{1,k\neq\tilde{k}}(\mathbf{x}_1)\mid{H}(\mathbf{x},t)\mid\psi(\mathbf{x},t)>dt=0
\end{eqnarray}
or
\begin{equation}
\setlength\abovedisplayskip{0pt}
\setlength\belowdisplayskip{0pt}
e^{\int_\tau^{\tau+T}\lambda_{k\neq\tilde{k}}^{-1}(\mathbf{x}_{1\textrm{-}},t)dt}=0.
\end{equation} 

If $\lambda_{k\neq\tilde{k}}^{-1}(\mathbf{x}_{1\textrm{-}},t)$ is determined by (11), then $\lambda_{k\neq\tilde{k}}^{-1}(\mathbf{x}_{1\textrm{-}},t)$ will depend on $H(\mathbf{x},t)\mid\psi(\mathbf{x},t)>$. Now consider a situation where no interaction exists between sub-system 1 and 1-. Due to the isolation, the two sub-systems will have their own Hamiltonians and wave functions. Denote $H_1(\mathbf{x}_1,t)$, $\mid\psi_1(\mathbf{x}_1,t)>$ and $H_{1\textrm{-}}(\mathbf{x}_{1\textrm{-}},t)$, $\mid\psi_{1\textrm{-}}(\mathbf{x}_{1\textrm{-}},t)>$ the Hamitonian and wave function for sub-system 1 and 1- respectively. The wave function $\mid\psi_{1\textrm{-}}(\mathbf{x}_{1\textrm{-}},t)>$ satisfies the equivalent of (1) or 
\begin{equation}
\setlength\abovedisplayskip{0pt}
\setlength\belowdisplayskip{0pt}
i\hbar\frac{\partial}{\partial{t}}\mid{\psi}_{1\textrm{-}}(\mathbf{x}_{1\textrm{-}},t)>=H_{1\textrm{-}}(\mathbf{x},t)\mid\psi_{1\textrm{-}}(\mathbf{x}_{1\textrm{-}},t)>
\end{equation} 
while $\mid\psi_1(\mathbf{x}_1,t)>$ satisfies the equivalent of (5) or 
{\setlength\arraycolsep{0pt}
\begin{eqnarray}
i\hbar\frac{\partial}{\partial{t}}\mid{\psi}_1(\mathbf{x}_1,t)> =&i\hbar\sum_k\lambda_{1,k}^{-1}(t)\mid\psi_{1,k}(\mathbf{x}_1)><\psi_{1,k}(\mathbf{x}_1)\mid\psi_1(\mathbf{x}_1,t)>\nonumber\\
&+H_1(\mathbf{x}_1,t)\mid\psi_1(\mathbf{x}_1,t)>
\end{eqnarray}
where $\lambda_{1,k}^{-1}(t)$ do not depend on sub-system 1-. On the other side, the system wave function $\mid\psi(\mathbf{x},t)>$ satisfies (5) with 
\begin{equation}
\setlength\abovedisplayskip{0pt}
\setlength\belowdisplayskip{0pt}
H(\mathbf{x},t)=H_1(\mathbf{x}_1,t)+H_{1\textrm{-}}(\mathbf{x}_{1\textrm{-}},t)
\end{equation} 
and
\begin{equation}
\setlength\abovedisplayskip{0pt}
\setlength\belowdisplayskip{0pt}
\mid\psi(\mathbf{x},t)>=\mid\psi_1(\mathbf{x}_1,t)>\mid\psi_{1\textrm{-}}(\mathbf{x}_{1\textrm{-}},t)>.
\end{equation} 
Due to the dependence of $\lambda_{k\neq\tilde{k}}^{-1}(\mathbf{x}_{1\textrm{-}},t)$ on $H(\mathbf{x},t)\mid\psi(\mathbf{x},t)>$, (5) together with (15) and (16) are not equivalent to (13) and (14). Therefore, the solution of (11) should be discarded.

Keeping the continuity and magnitude requirement of $\lambda_k^{-1}(\mathbf{x}_{1\textrm{-}},t)$ in mind, it is not difficult to verify that (12) requires the real part of $-\int_\tau^{\tau+T}\lambda_{k\neq\tilde{k}}^{-1}(\mathbf{x}_{1\textrm{-}},t)dt$ to approach infinity at $t=\tau+T$ or
\begin{equation}
\setlength\abovedisplayskip{0pt}
\setlength\belowdisplayskip{0pt}
\lim_{t \to \tau+T}\lambda_{k\neq\tilde{k}}^{-1}(\mathbf{x}_{1\textrm{-}},t)=\alpha_{k\neq\tilde{k}}(\mathbf{x}_{1\textrm{-}})(\tau+T-t)^{-\beta_{k\neq\tilde{k}}(\mathbf{x}_{1\textrm{-}})}
\end{equation} 
where $real(\alpha_{k\neq\tilde{k}}(\mathbf{x}_{1\textrm{-}})) < 0$, $|\alpha_{k\neq\tilde{k}}(\mathbf{x}_{1\textrm{-}})| < \infty$, and $\beta_{k\neq\tilde{k}}(\mathbf{x}_{1\textrm{-}}) \geq 1$. In this situation, although $e^{-\int_\tau^t\lambda_{k\neq\tilde{k}}^{-1}(\mathbf{x}_{1\textrm{-}},t)dt}$ blows up at $t=\tau+T$, since $<\psi_{1,k\neq\tilde{k}}(\mathbf{x}_1)\mid{H}(\mathbf{x},t)\mid\psi(\mathbf{x},t)>$ is finite, the right hand side of (10) will vanish at the end of the collapse. In other words, the solutions (17) of the conditions (12) are indeed the solutions of (9) subject to (6).

When $\beta_{k\neq\tilde{k}}(\mathbf{x}_{1\textrm{-}})=1$, $\lambda_{k\neq\tilde{k}}^{-1}(\mathbf{x}_{1\textrm{-}},\tau+T)<\psi_{1,k\neq\tilde{k}}(\mathbf{x}_1)\mid\psi(\mathbf{x},\tau+T)>$ is not necessarily zero. To see this, denote $\gamma_{k\neq\tilde{k}}(\mathbf{x}_{1\textrm{-}})$ the value of $<\psi_{1,k\neq\tilde{k}}(\mathbf{x}_1)\mid{H}(\mathbf{x},\tau+T)\mid\psi(\mathbf{x},\tau+T)>$, which represents a special case where $\frac{\partial}{\partial{t}}(H(\mathbf{x},\tau+T)\mid\psi(\mathbf{x},\tau+T)>)=0$. It is easy to verify that when $t$ approaches $\tau+T$, (9) has the following solution
\begin{equation}
\setlength\abovedisplayskip{0pt}
\setlength\belowdisplayskip{0pt}
i\hbar<\psi_{1,k}(\mathbf{x}_1)\mid\psi(\mathbf{x},t)>=-\frac{\gamma_k(\mathbf{x}_{1\textrm{-}})}{\alpha_k(\mathbf{x}_{1\textrm{-}})+1}(\tau+T-t).
\end{equation} 
Clearly as long as $\gamma_{k\neq\tilde{k}}(\mathbf{x}_{1\textrm{-}})\neq 0$, $\lambda_{k\neq\tilde{k}}^{-1}(\mathbf{x}_{1\textrm{-}},t)<\psi_{1,k\neq\tilde{k}}(\mathbf{x}_1)\mid\psi(\mathbf{x},\tau+T)>$ will not be zero. Since $\gamma_{k\neq\tilde{k}}(\mathbf{x}_{1\textrm{-}})=0$ is not guaranteed, in order to satisfy the continuity condition (8), $\beta_{k\neq\tilde{k}}(\mathbf{x}_{1\textrm{-}})$ should be strictly greater than one, which is equivalent to 
\begin{equation}
\setlength\abovedisplayskip{0pt}
\setlength\belowdisplayskip{0pt}
\frac{\partial}{\partial{t}}{\lambda}_{k\neq\tilde{k}}(\mathbf{x}_{1\textrm{-}},\tau+T)=0.
\end{equation} 

The value of $\lim_{t \to \tau+T}<\psi_{1,k\neq\tilde{k}}(\mathbf{x}_1)\mid{H}(\mathbf{x},t)\mid\psi(\mathbf{x},t)>$ can be generally denoted as $\gamma_{k\neq\tilde{k}}(\mathbf{x}_{1\textrm{-}})(\tau+T-t)^{\delta_{k\neq\tilde{k}}(\mathbf{x}_{1\textrm{-}})}$ where $\gamma_{k\neq\tilde{k}}(\mathbf{x}_{1\textrm{-}})\neq 0$ and $|\gamma_{k\neq\tilde{k}}(\mathbf{x}_{1\textrm{-}})|<\infty$. Compared to the magnitude of this value, the magnitude of the first term on the right hand side of (9) may be negligible, of the same order, or much larger. In case $i\hbar\lambda_k^{-1}(\mathbf{x}_{1\textrm{-}},\tau+T)<\psi_{1,k}(\mathbf{x}_1)\mid\psi(\mathbf{x},\tau+T)>$ is negligible, because of (9), $<\psi_{1,k}(\mathbf{x}_1)\mid\psi(\mathbf{x},\tau+T)>$ will be proportional to $(\tau+T-t)^{1+\delta_{k\neq\tilde{k}}}$. However, considering (17) with $\beta_{k\neq\tilde{k}}(\mathbf{x}_{1\textrm{-}})>1$, the magnitude of the first term on the right hand of (9) is actually much larger than that of the second term. The contradiction proves the incorrectness of this case. Similar analysis shows that the two terms cannot be the same order. So the only correct case is that the magnitude of $\gamma_{k\neq\tilde{k}}(\mathbf{x}_{1\textrm{-}})(\tau+T-t)^{\delta_{k\neq\tilde{k}}(\mathbf{x}_{1\textrm{-}})}$ is negligible when compared to that of $\lambda_{k\neq\tilde{k}}^{-1}(\mathbf{x}_{1\textrm{-}},t)\nu_{k\neq\tilde{k}}(\mathbf{x}_{1\textrm{-}},t)(\tau+T-t)^{1+\delta_{k\neq\tilde{k}}(\mathbf{x}_{1\textrm{-}})}$. In other words, toward the end of the collapse, (9) reduces to 
\begin{equation}
\setlength\abovedisplayskip{0pt}
\setlength\belowdisplayskip{0pt}
\frac{\partial}{\partial{t}}<\psi_{1,k\neq\tilde{k}}(\mathbf{x}_1)\mid{\psi}(\mathbf{x},t)>=\lambda_{k\neq\tilde{k}}^{-1}(\mathbf{x}_{1\textrm{-}},t)<\psi_{1,k\neq\tilde{k}}(\mathbf{x}_1)\mid\psi(\mathbf{x},t)>.
\end{equation}

Eq.(20) defines the asymptotic behavior of $<\psi_{1,k\neq\tilde{k}}(\mathbf{x}_1)\mid\psi(\mathbf{x},t)>$, i.e., $<\psi_{1,k\neq\tilde{k}}(\mathbf{x}_1)\mid\psi(\mathbf{x},t)>$ being proportional to  $e^{\int_\tau^t\lambda_{k\neq\tilde{k}}^{-1}(\mathbf{x}_{1\textrm{-}},t)dt}$ vanishes when $t$ approaches $\tau+T$. Keeping (17) in mind, the asymptotic behavior can be expressed as 
\begin{equation}
\setlength\abovedisplayskip{0pt}
\setlength\belowdisplayskip{0pt}
\lim_{t \to \tau+T}<\psi_{1,k\neq\tilde{k}}(\mathbf{x}_1)\mid\psi(\mathbf{x},t)>\propto
{e^{\frac{\alpha_{k\neq\tilde{k}}(\mathbf{x}_{1\textrm{-}})}{\beta_{k\neq\tilde{k}}(\mathbf{x}_{1\textrm{-}})-1}(\tau+T-t)^{1-\beta_{k\neq\tilde{k}}(\mathbf{x}_{1\textrm{-}})}}}.
\end{equation} 
Since the magnitude of $<\psi_{1,k\neq\tilde{k}}(\mathbf{x}_1)\mid{H}(\mathbf{x},\tau+T)\mid\psi(\mathbf{x},\tau+T)>$ is less than the magnitude of $\lambda_{k\neq\tilde{k}}^{-1}(\mathbf{x}_{1\textrm{-}},\tau+T)<\psi_{1,k\neq\tilde{k}}(\mathbf{x}_1)\mid\psi(\mathbf{x},\tau+T)>$,
\begin{equation}
\setlength\abovedisplayskip{3pt}
\setlength\belowdisplayskip{0pt}
<\psi_{1,k\neq\tilde{k}}(\mathbf{x}_1)\mid{H}(\mathbf{x},\tau+T)\mid\psi(\mathbf{x},\tau+T)>=0.
\end{equation}

As for the term $\lambda_{\tilde{k}}^{-1}(\mathbf{x}_{1\textrm{-}},\tau+T)\mid\psi_{1,{\tilde{k}}}(\mathbf{x}_1)><\psi_{1,{\tilde{k}}}(\mathbf{x}_1)\mid\psi(\mathbf{x},\tau+T)>$ in the continuity condition (8), it follows from $<\psi_{1,\tilde{k}}(\mathbf{x}_1)\mid\psi(\mathbf{x},\tau+T)>\neq 0$ that 
\begin{equation}
\setlength\abovedisplayskip{3pt}
\setlength\belowdisplayskip{0pt}
\lambda_{\tilde{k}}^{-1}(\mathbf{x}_{1\textrm{-}},\tau+T)=0.
\end{equation}

\section{The revelation of the preferred energy eigenfunctions}

\subsection{The restriction on the preferred-basis of isolated sub-system}

At the end of the collapse, the wave function satisfies (6), (22), and (23). Equations (22) require appropriate functions as the preferred-basis. A particular case will be the collapse of sub-system 1 that does not interact with the rest of the system. Making use of (15) and (16), it follows from (6) that the end state $\mid\psi_1(\mathbf{x}_1,\tau+T)>$ is one of the basis function $\mid\psi_{1,\tilde{k}}(\mathbf{x}_1)>$ scaled by a coefficient of magnitude one. So the orthogonal conditions (22) become
\begin{equation}
\setlength\abovedisplayskip{3pt}
\setlength\belowdisplayskip{0pt}
<\psi_{1,k\neq\tilde{k}}(\mathbf{x}_1)\mid{H}_1(\mathbf{x}_1,\tau+T)\mid\psi_{1,k}(\mathbf{x}_1)>=0.
\end{equation}
Due to the completeness of the preferred-basis, it follows from (24) that $H_1(\mathbf{x}_1,\tau+T)\mid\psi_{1,\tilde{k}}(\mathbf{x}_1)>$ can only contain the basis function $\mid\psi_{1,\tilde{k}}(\mathbf{x}_1)>$, i.e., 
\begin{equation}
\setlength\abovedisplayskip{0pt}
\setlength\belowdisplayskip{0pt}
H_1(\mathbf{x}_1,\tau+T)\mid\psi_{1,\tilde{k}}(\mathbf{x}_1)>=\epsilon_{1,\tilde{k}}\mid\psi_{1,\tilde{k}}(\mathbf{x}_1)>
\end{equation}
with some constant $\epsilon_{1,\tilde{k}}$. This equation clearly shows that the preferred-basis of an isolated system is the set of eigenfunctions of the Hamiltonian at the end of the collapse.

In the standard quantum mechanics, the weights of preferred-basis functions in the expansion of the system wave function determine the probabilities to collapse to the basis functions. That's to say, preferred-basis does not change with the weights of basis functions. Such requirement can be used to relate the basis functions $\mid\psi_{1,\tilde{k}}(\mathbf{x}_1)>$ to $H_1(\mathbf{x}_1,\tau)$. However, to make the analysis self-contained, it is explicitly assumed that any set of determination equations of the preferred-basis functions is independent of the basis functions' weights in the system wave function, which is the first assumption made in this paper on the determination of preferred-basis.

According to the assumption, given the preferred-basis functions $\mid\psi_{1,k}(\mathbf{x}_1)>$, the weights of the basis functions prior the collapse can be arbitrarily set as long as the system wave function is normalized. So cases can be made such that the sub-system wave function $\mid\psi_1(\mathbf{x}_1,\tau)>$ at the beginning of the collapse only contains one basis function. Consequently the ending Hamiltonian will be the same as that at the beginning of the collapse as long the latter is stationary. Therefore for stationary Hamiltonian $H_1(\mathbf{x}_1,\tau)$, $\mid\psi_{1,k}(\mathbf{x}_1)>$ have to be the set of energy eigenfunctions. 

\subsection{No extra restriction on the preferred-basis}

More often interaction exists between sub-system 1 and 1-. In order for (6) and (22) to restrict the preferred-basis, during the collapse process the system Hamiltonian $H(\mathbf{x},t)$ should be stationary and $<\psi_{1,k}(\mathbf{x}_1)\mid\psi(\mathbf{x},t)>$ should be proportional to $<\psi_{1,k}(\mathbf{x}_1)\mid\psi(\mathbf{x},\tau)>$. Otherwise it is the change in $H(\mathbf{x},t)$ and $<\psi_{1,k}(\mathbf{x}_1)\mid\psi(\mathbf{x},t)>$ that make (22) to hold. That means $\mid\psi(\mathbf{x},\tau)>$ only contains one basis function, e.g., $\mid\psi_{1,\tilde{k}}(\mathbf{x}_1)><\psi_{1,\tilde{k}}(\mathbf{x}_1)\mid\psi(\mathbf{x},\tau)>$ and 
 \begin{equation}
\setlength\abovedisplayskip{3pt}
\setlength\belowdisplayskip{0pt}
<\psi_{1,k\neq\tilde{k}}(\mathbf{x}_1)\mid{H}(\mathbf{x},\tau)\mid\psi_{1,\tilde{k}}(\mathbf{x}_1)>\mid\phi_{1\textrm{-},\tilde{k}}(\mathbf{x}_{1\textrm{-}})>=0
\end{equation}
where $\mid\phi_{1\textrm{-},\tilde{k}}(\mathbf{x}_{1\textrm{-}})>=<\psi_{1,\tilde{k}}(\mathbf{x}_1)\mid\psi(\mathbf{x},\tau)>$. The system Hamiltonian $H(\mathbf{x},\tau)$ can be split into three components---the Hamiltonian $H_1(\mathbf{x}_1,\tau)$ of sub-system 1, the Hamiltonian $H_{1\textrm{-}}(\mathbf{x}_{1\textrm{-}},\tau)$ of sub-system 1-, and the interaction term $H_{1,1\textrm{-}}(\mathbf{x}_1,\mathbf{x}_{1\textrm{-}},\tau)$ that entangles the two sub-systems and contains algebraic functions of $\mathbf{x}_1$ and $\mathbf{x}_{1\textrm{-}}$.

The completeness of the preferred-basis  $\mid\psi_{1,k}(\mathbf{x}_1)>$ together with (26) forces $H(\mathbf{x},\tau)\mid\psi_{1,\tilde{k}}(\mathbf{x}_1)>\mid\phi_{1\textrm{-},\tilde{k}}(\mathbf{x}_{1\textrm{-}})>$ to be the product of $\mid\psi_{1,\tilde{k}}(\mathbf{x}_1)>$ and some function in $\mathbf{x}_{1\textrm{-}}$ domain. Since $H_{1\textrm{-}}(\mathbf{x}_{1\textrm{-}},\tau)$ only acts on sub-system 1-, $(H_1(\mathbf{x}_1,\tau)+H_{1,1\textrm{-}}(\mathbf{x}_1,\mathbf{x}_{1\textrm{-}},\tau))\mid\psi_{1,\tilde{k}}(\mathbf{x}_1)>\mid\phi_{1\textrm{-},\tilde{k}}(\mathbf{x}_{1\textrm{-}})>$ should also be a product of $\mid\psi_{1,\tilde{k}}(\mathbf{x}_1)>$ and some function of $\mathbf{x}_{1\textrm{-}}$. Denoting such function by $\mid\varphi_{1\textrm{-},\tilde{k}}(\mathbf{x}_{1\textrm{-}})>$, the following equation is obtained
\setlength\abovedisplayskip{0pt}
\setlength\belowdisplayskip{0pt}
\begin{eqnarray}
(H_1(\mathbf{x}_1,\tau)+H_{1,1\textrm{-}}(\mathbf{x}_1,\mathbf{x}_{1\textrm{-}},\tau))\mid\psi_{1,\tilde{k}}(\mathbf{x}_1)>\mid\phi_{1\textrm{-},\tilde{k}}(\mathbf{x}_{1\textrm{-}})>\nonumber\\
=\mid\psi_{1,\tilde{k}}(\mathbf{x}_1)>\mid\varphi_{1\textrm{-},\tilde{k}}(\mathbf{x}_{1\textrm{-}})>.
\end{eqnarray}

In view of the algebraic functions in $H_{1,1\textrm{-}}(\mathbf{x}_1,\mathbf{x}_{1\textrm{-}},\tau)$, for any two points $\mathbf{x}_{1\textrm{-}}^{p1}$ and $\mathbf{x}_{1\textrm{-}}^{p2}$ in the $\mathbf{x}_{1\textrm{-}}$ domain, the instantiations of (27) will be 
\setlength\abovedisplayskip{0pt}
\setlength\belowdisplayskip{0pt}
\begin{eqnarray}
(H_1(\mathbf{x}_1,\tau)+H_{1,1\textrm{-}}(\mathbf{x}_1,\mathbf{x}_{1\textrm{-}}^{p1},\tau))\mid\psi_{1,\tilde{k}}(\mathbf{x}_1)>\mid\phi_{1\textrm{-},\tilde{k}}(\mathbf{x}_{1\textrm{-}}^{p1})>=\nonumber\\
\mid\psi_{1,\tilde{k}}(\mathbf{x}_1)>\mid\varphi_{1\textrm{-},\tilde{k}}(\mathbf{x}_{1\textrm{-}}^{p1})>
\end{eqnarray}
and
\setlength\abovedisplayskip{0pt}
\setlength\belowdisplayskip{0pt}
\begin{eqnarray}
(H_1(\mathbf{x}_1,\tau)+H_{1,1\textrm{-}}(\mathbf{x}_1,\mathbf{x}_{1\textrm{-}}^{p2},\tau))\mid\psi_{1,\tilde{k}}(\mathbf{x}_1)>\mid\phi_{1\textrm{-},\tilde{k}}(\mathbf{x}_{1\textrm{-}}^{p2})>\nonumber\\
=\mid\psi_{1,\tilde{k}}(\mathbf{x}_1)>\mid\varphi_{1\textrm{-},\tilde{k}}(\mathbf{x}_{1\textrm{-}}^{p2})>.
\end{eqnarray}
When $\mid\phi_{1\textrm{-},\tilde{k}}(\mathbf{x}_{1\textrm{-}}^{p1})>$ and $\mid\phi_{1\textrm{-},\tilde{k}}(\mathbf{x}_{1\textrm{-}}^{p2})>$ are not zero, dividing both sides of (28) and (29) by $\mid\phi_{1\textrm{-},\tilde{k}}(\mathbf{x}_{1\textrm{-}}^{p1})>$ and $\mid\phi_{1\textrm{-},\tilde{k}}(\mathbf{x}_{1\textrm{-}}^{p2})>$ respectively and subtracting one resultant equation from the other yields
\setlength\abovedisplayskip{0pt}
\setlength\belowdisplayskip{0pt}
\begin{eqnarray}
(H_{1,1\textrm{-}}(\mathbf{x}_1,\mathbf{x}_{1\textrm{-}}^{p1},\tau)-H_{1,1\textrm{-}}(\mathbf{x}_1,\mathbf{x}_{1\textrm{-}}^{p2},\tau))\mid\psi_{1,\tilde{k}}(\mathbf{x}_1)>=\mid\psi_{1,\tilde{k}}(\mathbf{x}_1)>\nonumber\\
(\mid\varphi_{1\textrm{-},\tilde{k}}(\mathbf{x}_{1\textrm{-}}^{p1})>\mid\phi_{1\textrm{-},\tilde{k}}^{-1}(\mathbf{x}_{1\textrm{-}}^{p1})>-\mid\varphi_{1\textrm{-},\tilde{k}}(\mathbf{x}_{1\textrm{-}}^{p2})>\mid\phi_{1\textrm{-},\tilde{k}}^{-1}(\mathbf{x}_{1\textrm{-}}^{p2})>).
\end{eqnarray}
Since $H_{1,1\textrm{-}}(\mathbf{x}_1,\mathbf{x}_{1\textrm{-}}^{p1},\tau)-H_{1,1\textrm{-}}(\mathbf{x}_1,\mathbf{x}_{1\textrm{-}}^{p2},\tau))$ is also an algrebraic function of $\mathbf{x}_1$, (30) forces $\mid\psi_{1,\tilde{k}}(\mathbf{x}_1)>$ to be a position eigenfunction.

However, $H(\mathbf{x},\tau)$ contains second order derivative on position. When $\mid\psi_{1,\tilde{k}}(\mathbf{x}_1)>$ is a position eigenfunction, $H(\mathbf{x},\tau)\mid\psi_{1,\tilde{k}}(\mathbf{x}_1)><\psi_{1,\tilde{k}}(\mathbf{x}_1)\mid\psi(\mathbf{x},\tau)>$ cannot be proportional to $\mid\psi_{1,\tilde{k}}(\mathbf{x}_1,\tau)>$, which contradicts (26). Therefore the orthogonal conditions (22) do not restrict the preferred-basis for the general system. 

\section{The determination equations of preferred-basis}
As proved in Section 3, the collapse equation conditionally requires the preferred-basis of isolated sub-system to be the energy eigenfunctions. Those special cases can be used to find the determination equations of preferred-basis for any kind of sub-system 1. Since both Schr\"odinger equation and the collapse equation contain the whole system Hamiltonian, it is reasonable to assume that the determination equations depend on the whole system Hamiltonian, not part of it, which is the second assumption made in this paper on the determination of preferred-basis.

\subsection{The criteria of good determination equations}

Functions $\mid\psi_{1,k}(\mathbf{x}_1)><\psi_{1,k}(\mathbf{x}_1)\mid\psi_1(\mathbf{x}_1,\tau)>$ being the eigenfunctions of $H_1(\mathbf{x}_1,\tau)$ has two meanings. First, $H_1(\mathbf{x}_1,\tau)\mid\psi_{1,k}(\mathbf{x}_1)><\psi_{1,k}(\mathbf{x}_1)\mid\psi_1(\mathbf{x}_1,\tau)>$ is proportional to $\mid\psi_{1,k}(\mathbf{x}_1)><\psi_{1,k}(\mathbf{x}_1)\mid\psi_1(\mathbf{x}_1,\tau)>$. Second, $H_1(\mathbf{x}_1,\tau)\mid\psi_{1,k}(\mathbf{x}_1)><\psi_{1,k}(\mathbf{x}_1)\mid\psi_1(\mathbf{x}_1,\tau)>$ is orthogonal to all $\mid\psi_{1,k'\neq{k}}(\mathbf{x}_1)><\psi_{1,k'\neq{k}}(\mathbf{x}_1)\mid\psi_1(\mathbf{x}_1,\tau)>$. Of course, the mathematical description of each meaning is not unique. For examples, equations $H_1(\mathbf{x}_1,\tau)\mid\psi_{1,k}(\mathbf{x}_1)><\psi_{1,k}(\mathbf{x}_1)\mid\psi_1(\mathbf{x}_1,\tau)>=\epsilon_{1,k}\mid\psi_{1,k}(\mathbf{x}_1)><\psi_{1,k}(\mathbf{x}_1)\mid\psi_1(\mathbf{x}_1,\tau)>$ and $H^2_1(\mathbf{x}_1,\tau)\mid\psi_{1,k}(\mathbf{x}_1)><\psi_{1,k}(\mathbf{x}_1)\mid\psi_1(\mathbf{x}_1,\tau)>=\epsilon
^2_{1,k}\mid\psi_{1,k}(\mathbf{x}_1)><\psi_{1,k}(\mathbf{x}_1)\mid\psi_1(\mathbf{x}_1,\tau)>$ equivalently describe the first meaning. But no matter how the expression is, it should contain some function of $H_1(\mathbf{x}_1,\tau)$ operating on $\mid\psi_{1,k}(\mathbf{x}_1)><\psi_{1,k}(\mathbf{x}_1)\mid\psi_1(\mathbf{x}_1,\tau)>$. The good expression should generate consistent results after its special Hamiltonian and wave functions are replaced by their generic counterparts. Clearly such expression meets the requirement in the second assumption. To align with the first assumption, the normalized wave function $\mid\psi_{1,k}(\mathbf{x}_1)>\bm{<}\psi_{1,k}(\mathbf{x}_1)\mid\psi_1(\mathbf{x}_1,\tau)\bm{>}$ should be used rather than  $\mid\psi_{1,k}(\mathbf{x}_1)><\psi_{1,k}(\mathbf{x}_1)\mid\psi_1(\mathbf{x}_1,\tau)>$. Here $\bm{<}{\cdots}\bm{>}$ denotes the normalization of $<{\cdots}>$. Using notation $||\cdots||$ to denote norm, e.g., $||<\psi_{1,k}(\mathbf{x}_1)\mid\psi_1(\mathbf{x}_1,\tau)>|| = \sqrt{<\hat{\phi}_{1\textrm{-},k}(\mathbf{x}_{1\textrm{-}})\mid\hat{\phi}_{1\textrm{-},k}(\mathbf{x}_{1\textrm{-}})>}$ where $\mid\hat{\phi}_{1\textrm{-},k}(\mathbf{x}_{1\textrm{-}})>=<\psi_{1,k}(\mathbf{x}_1)\mid\psi_1(\mathbf{x}_1,\tau)>$, then $\bm{<}\psi_{1,k}(\mathbf{x}_1)\mid\psi_1(\mathbf{x}_1,\tau)\bm{>}=||<\psi_{1,k}(\mathbf{x}_1)\mid\psi_1(\mathbf{x}_1,\tau)>||^{-1}<\psi_{1,k}(\mathbf{x}_1)\mid\psi_1(\mathbf{x}_1,\tau)>$. 

Now consider the whole system that consists of non-interacting sub-system 1 and 1-. If the original equation contains $H^2_1(\mathbf{x}_1,\tau)\mid\psi_{1,k}(\mathbf{x}_1)>\bm{<}\psi_{1,k}(\mathbf{x}_1)\mid\psi_1(\mathbf{x}_1,\tau)\bm{>}$, then the generalized equation will contain $\mid\psi_{1\textrm{-}}(\mathbf{x}_{1\textrm{-}},\tau)>H^2_1(\mathbf{x}_1,\tau)\mid\psi_{1,k}(\mathbf{x}_1)>\bm{<}\psi_{1,k}(\mathbf{x}_1)\mid\psi_1(\mathbf{x}_1,\tau)\bm{>}+2H_1(\mathbf{x}_1,\tau)\mid\psi_{1,k}(\mathbf{x}_1)>\bm{<}\psi_{1,k}(\mathbf{x}_1)\mid\psi_1(\mathbf{x}_1,\tau)\bm{>}H_{1\textrm{-}}(\mathbf{x}_{1\textrm{-}},\tau)\mid\psi_{1\textrm{-}}(\mathbf{x}_{1\textrm{-}},\tau)>+\mid\psi_{1,k}(\mathbf{x}_1)>\bm{<}\psi_{1,k}(\mathbf{x}_1)\mid\psi_1(\mathbf{x}_1,\tau)\bm{>}H^2_{1\textrm{-}}(\mathbf{x}_{1\textrm{-}},\tau)\mid\psi_{1\textrm{-}}(\mathbf{x}_{1\textrm{-}},\tau)>$. Due to the arbitrariness of $\mid\psi_{1\textrm{-}}(\mathbf{x}_{1\textrm{-}},\tau)>$ or $H_{1\textrm{-}}(\mathbf{x}_{1\textrm{-}},\tau)$, no matter what arithmetic operation is applied, the generalized equation will depend on sub-system 1-. In other words, the inclusion of an isolated sub-system changes the preferred-basis of the isolated sub-sytem 1. To avoid this inconsistence, $H^2_1(\mathbf{x}_1,\tau)$ cannot exist in the original equations. By similar argument, it can be proved that no other nonlinear function of $H_1(\mathbf{x}_1,\tau)$ can exist in the original equations either.

The linear term $H_1(\mathbf{x}_1,\tau)\mid\psi_{1,k}(\mathbf{x}_1)>\bm{<}\psi_{1,k}(\mathbf{x}_1)\mid\psi_1(\mathbf{x}_1,\tau)\bm{>}$ becomes $\mid\psi_{1\textrm{-}}(\mathbf{x}_{1\textrm{-}},\tau)>H_1(\mathbf{x}_1,\tau)\mid\psi_{1,k}(\mathbf{x}_1)>\bm{<}\psi_{1,k}(\mathbf{x}_1)\mid\psi_1(\mathbf{x}_1,\tau)\bm{>}+\mid\psi_{1,k}(\mathbf{x}_1)>\bm{<}\psi_{1,k}(\mathbf{x}_1)\mid\psi_1(\mathbf{x}_1,\tau)\bm{>}H_{1\textrm{-}}(\mathbf{x}_{1\textrm{-}},\tau)\mid\psi_{1\textrm{-}}(\mathbf{x}_{1\textrm{-}},\tau)>$ in the enlarged system. In order to restore the original description of the energy eigenfunctions, some arithmetic operation needs to be carried out in the $\mathbf{x}_{1\textrm{-}}$ or $\mathbf{x}$ domain depending on which meaning of the eigenfunctions is used.

In a more general situation the two sub-systems interact with each other. Now the preferred-basis of sub-system 1 may change to other set of functions. But given any complete set of orthonormal functions, the to-be-determined preferred-basis functions can be written as linear combinations of the given functions. For the convenience of analysis without loss of generality, assume the given set consists of $N_1$ functions. So there are $N_1$ preferred-basis functions and each function contains $N_1$ unknown coefficients in its expression as a linear combination of the given basis functions. The total number of unknowns will be $N^2_1$. Meanwhile constraints are needed to guarantee the orthonormalization of the preferred-basis. It is easy to see that $\frac{N_1(N_1+1)}{2}$ constraints are needed. So the good original equations are those such that their generalization yields $\frac{N_1(N_1-1)}{2}$ more new constraints than the new unknowns.

\subsection{The valid determination equations}
when the original equations describe the first meaning of the energy eigenfunctions, the generalized term $H(\mathbf{x},\tau)\mid\psi_{1,k}(\mathbf{x}_1)>\bm{<}\psi_{1,k}(\mathbf{x}_1)\mid\psi(\mathbf{x},\tau)\bm{>}$ can only be proportional to $E_k\mid\psi_{1,k}(\mathbf{x}_1)>\bm{<}\psi_{1,k}(\mathbf{x}_1)\mid\psi(\mathbf{x},\tau)\bm{>}$ after the application of whatever arithmetic operation in the $\mathbf{x}_{1\textrm{-}}$ domain defined in the original equations. But generally $H(\mathbf{x},\tau)\mid\psi_{1,k}(\mathbf{x}_1)>\bm{<}\psi_{1,k}(\mathbf{x}_1)\mid\psi(\mathbf{x},\tau)\bm{>}$ contains terms proportional to all functions $\mid\psi_{1,k'}(\mathbf{x}_1)>$. Therefore each term $H(\mathbf{x},\tau)\mid\psi_{1,k}(\mathbf{x}_1)>\bm{<}\psi_{1,k}(\mathbf{x}_1)\mid\psi(\mathbf{x},\tau)\bm{>}$ will result in $N_1$ constraints and one unknown variable $E_k$. Since the original functions have to contain $H_1(\mathbf{x},\tau)\mid\psi_{1,k}(\mathbf{x}_1)>\bm{<}\psi_{1,k}(\mathbf{x}_1)\mid\psi_1(\mathbf{x}_1,\tau)\bm{>}$ for all $k$, the total number of constraints and unknowns will be $N^2_1$ and $N_1$ respectively. Clearly the difference between the constraints and the unknowns is greater than $\frac{N_1(N_1-1)}{2}$ for $N_1>1$, so generally there is no solution for the preferred-basis. In other words, the generalized equations are not valid.

To describe the second meaning of the energy eigenfunctions, the inner products in the $\mathbf{x}$ domain have to be used. One version of the original equations are
\setlength\abovedisplayskip{0pt}
\setlength\belowdisplayskip{0pt}
\begin{eqnarray}
\fl \bm{<}\psi_1(\mathbf{x}_1,\tau)\mid\psi_{1,k'}(\mathbf{x}_1)\bm{>}<\psi_{1,k'}(\mathbf{x}_1)\mid{H}(\mathbf{x},\tau)\mid\psi_{1,k}(\mathbf{x}_1)>\bm{<}\psi_{1,k}(\mathbf{x}_1)\mid\psi_1(\mathbf{x}_1,\tau)\bm{>}\nonumber\\
\fl =\epsilon_{1,k}\delta_{k'k} \qquad \textrm{for} \quad k'\geq{k}
\end{eqnarray}
where $\delta_{k'k}$ is the Kronecker delta. The generalization is 
\setlength\abovedisplayskip{0pt}
\setlength\belowdisplayskip{0pt}
\begin{eqnarray}
\fl \bm{<}\psi(\mathbf{x},\tau)\mid\psi_{1,k'}(\mathbf{x}_1)\bm{>}<\psi_{1,k'}(\mathbf{x}_1)\mid{H}(\mathbf{x},\tau)\mid\psi_{1,k}(\mathbf{x}_1)>\bm{<}\psi_{1,k}(\mathbf{x}_1)\mid\psi(\mathbf{x},\tau)\bm{>}\nonumber\\
\fl =E_{1,k}\delta_{k'k}  \qquad \textrm{for} \quad k'\geq{k},
\end{eqnarray}
which yield $\frac{N_1(N_1+1)}{2}$ constraints and $N_1$ unknowns $E_{1,k}$. Because the difference is exactly $\frac{N_1(N_1-1)}{2}$, the group of equations (32) are good determination equations.

The only variance in the original equations is to use linear combinations of (31) to replace (31). Correspondingly the generalized equations will undergo the same linear transformation. As a result, they will be equivalent to (32).

Combining the mathematical descriptions of the two meanings of the original energy eigenfunctions does not lead to the correct number of constraints and unknowns. For example, if one basis function is defined on the first meaning and the rest on the second meaning, then the number of constraints and unknowns will be $N_1+\frac{N_1(N_1+1)}{2}-1$ and $N_1$. The difference is greater than $\frac{N_1(N_1-1)}{2}$ for $N_1>1$.
 
Therefore only (32) yield the same number of constraints and unknowns. They are the only valid set of equations that determine the preferred-basis in general situations.

\subsection{A few asymptotic solutions}
It is generally difficult to find the exact solutions from (32). Luckily some asymptotic solutions can be easily obtained. In fact, if a sub-system weakly interacts with others, then (32) are approximately energy eigen-equations. So the preferred-basis is close to the set of energy eigenfunctions of the sub-system Hamiltonian. This explains why nearly isolated atoms generate spectrum lines corresponding to the difference in the energy eigenvalues. In case the sub-system strongly interact with others by algebraic functions of position in the system Hamiltonian, then the Hamiltonian $H(\mathbf{x},\tau)$ in (30) can be approximately replaced with the algebraic functions. In this situation the solutions of (32) are nothing but quasi-position eigenfunctions, which explains the localization in the double-slit experiments, etc.

Another interesting case will be a sub-system that partially interacts with the environment by strong algebraic functions of position in the system Hamiltonian. Clearly for the part not interacting with the environment, the basis functions are the energy eigenfunctions of the sub-system. The interaction will result in quasi-position eigenfunctions as the preferred-basis for the interacting part of the sub-system. So the overall basis will be the combination of energy eigenfunctions and quasi-position eigenfunctions. This type of basis satisfactorily describes the behavior of macroscopic objects, e.g., semiconductor devices [20], that have well defined macroscopic location but meanwhile their internal objects only transition between energy eigenstates.

\section{The cyclic functions in the collapse equation}

Plugging (17) and (21) into the right hand side of (20) yields the following asymptotic expression of $\lambda_{k\neq\tilde{k}}^{-1}(\mathbf{x}_{1\textrm{-}},t)<\psi_{1,k\neq\tilde{k}}(\mathbf{x}_1)\mid\psi(\mathbf{x},t)>$
\setlength\abovedisplayskip{0pt}
\setlength\belowdisplayskip{0pt}
\begin{eqnarray}
\lim_{t \to \tau+T}\lambda_{k\neq\tilde{k}}^{-1}(\mathbf{x}_{1\textrm{-}},t)<\psi_{1,k\neq\tilde{k}}(\mathbf{x}_1)\mid\psi(\mathbf{x},t)>\propto\nonumber\\
{(\tau+T-t)^{-\beta_{k\neq\tilde{k}}(\mathbf{x}_{1\textrm{-}})}e^{\frac{\alpha_{k\neq\tilde{k}}(\mathbf{x}_{1\textrm{-}})}{\beta_{k\neq\tilde{k}}(\mathbf{x}_{1\textrm{-}})-1}(\tau+T-t)^{1-\beta_{k\neq\tilde{k}}(\mathbf{x}_{1\textrm{-}})}}}.
\end{eqnarray}
Because $e^{\frac{\alpha_{k\neq\tilde{k}}(\mathbf{x}_{1\textrm{-}})}{\beta_{k\neq\tilde{k}}(\mathbf{x}_{1\textrm{-}})-1}(\tau+T-t)^{1-\beta_{k\neq\tilde{k}}(\mathbf{x}_{1\textrm{-}})}}$ decays rapidly, the first order time derivative of $\lambda_{k\neq\tilde{k}}^{-1}(\mathbf{x}_{1\textrm{-}},t)<\psi_{1,k\neq\tilde{k}}(\mathbf{x}_1)\mid\psi(\mathbf{x},t)>$ vanishes toward the end of the collapse. Making use of this zero value, the derivative of (5) at $t=\tau+T$ reduces to
\setlength\abovedisplayskip{0pt}
\setlength\belowdisplayskip{0pt}
\begin{eqnarray}
\fl i\hbar\frac{\partial^2}{\partial{t}^2}\mid{\psi}(\mathbf{x},t)>=&\frac{\partial}{\partial{t}}(\lambda_{\tilde{k}}^{-1}(\mathbf{x}_{1\textrm{-}},t)\mid\psi_{1,\tilde{k}}(\mathbf{x}_1)><\psi_{1,\tilde{k}}(\mathbf{x}_1)\mid\psi(\mathbf{x},t)>\nonumber\\
&+H(\mathbf{x},t)\mid\psi(\mathbf{x},t)>).
\end{eqnarray}
Because of (23) and the finiteness of $\frac{\partial}{\partial{t}}(<\psi_{1,\tilde{k}}(\mathbf{x}_1)\mid\psi(\mathbf{x},t)>)$, which is guaranteed by (5), under the condition
\begin{equation}
\setlength\abovedisplayskip{0pt}
\setlength\belowdisplayskip{0pt}
\frac{\partial}{\partial{t}}{\lambda}^{-1}_{\tilde{k}}(\mathbf{x}_{1\textrm{-}},\tau+T)=0
\end{equation} 
and that $\frac{\partial}{\partial{t}}H(\mathbf{x},t)$ is continuous at $t=\tau+T$, (34) further simplifies to 
\begin{equation}
\setlength\abovedisplayskip{0pt}
\setlength\belowdisplayskip{0pt}
i\hbar\frac{\partial^2}{\partial{t}^2}\mid{\psi}(\mathbf{x},\tau+T)>=\frac{\partial}{\partial{t}}(H(\mathbf{x},\tau+T)\mid\psi(\mathbf{x},\tau+T)>),
\end{equation} 
which means the second order time derivative of $\mid{\psi}(\mathbf{x},t)>$ is continuous at the end of the collapse.

It will be interesting to require the second order time derivative of the wave function to be continuous at the beginning of the collapse. Again under the condition that $\frac{\partial}{\partial{t}}H(\mathbf{x},t)$ is continuous at $t=\tau$, $\lambda_k^{-1}(\mathbf{x}_{1\textrm{-}},t)$ needs to satisfy 
\begin{equation}
\setlength\abovedisplayskip{0pt}
\setlength\belowdisplayskip{0pt}
\frac{\partial}{\partial{t}}(\lambda_k^{-1}(\mathbf{x}_{1\textrm{-}},\tau)<\psi_{1,k}(\mathbf{x}_1)\mid\psi(\mathbf{x},\tau)>)=0
\end{equation} 
or
\begin{equation}
\setlength\abovedisplayskip{0pt}
\setlength\belowdisplayskip{0pt}
\frac{\partial}{\partial{t}}{\lambda_k^{-1}}(\mathbf{x}_{1\textrm{-}},\tau)=0.
\end{equation} 
for general situation where $<\psi_{1,k}(\mathbf{x}_1)\mid\psi(\mathbf{x},\tau)>\neq 0$. 

For $k\neq{\tilde{k}}$, (17), (19), and (38) are equivalent to 
\begin{equation}
\setlength\abovedisplayskip{0pt}
\setlength\belowdisplayskip{0pt}
\lambda_{k\neq{\tilde{k}}}^{-1}(\mathbf{x}_{1\textrm{-}},t)=\frac{f_{k\neq{\tilde{k}}}(\mathbf{x}_{1\textrm{-}},t)-1}{f_{k\neq{\tilde{k}}}(\mathbf{x}_{1\textrm{-}},t)}
\end{equation} 
where $f_{k\neq{\tilde{k}}}(\mathbf{x}_{1\textrm{-}},t)$ satisfies the following conditions
\begin{equation}
\setlength\abovedisplayskip{0pt}
\setlength\belowdisplayskip{0pt}
f_{k\neq{\tilde{k}}}(\mathbf{x}_{1\textrm{-}},\tau)=1,
\end{equation} 
\begin{equation}
\setlength\abovedisplayskip{0pt}
\setlength\belowdisplayskip{0pt}
\frac{\partial}{\partial{t}}f_{k\neq{\tilde{k}}}(\mathbf{x}_{1\textrm{-}},\tau)=0,
\end{equation} 
\begin{equation}
\setlength\abovedisplayskip{0pt}
\setlength\belowdisplayskip{0pt}
f_{k\neq{\tilde{k}}}(\mathbf{x}_{1\textrm{-}},\tau+T)=0,
\end{equation} 
and
\begin{equation}
\setlength\abovedisplayskip{0pt}
\setlength\belowdisplayskip{0pt}
\frac{\partial}{\partial{t}}f_{k\neq{\tilde{k}}}(\mathbf{x}_{1\textrm{-}},\tau+T)=0.
\end{equation} 

According to Sturm-Liouville theory [19], the solution of a 2\textsuperscript{nd} order differential equation $\frac{d^2y(t)}{dt^2}+\omega^2y(t)=0$ is a complete set of basis functions if $y(\tau)=0$ and $y(\tau+T)=0$ for all possible $\omega$. The general solutions are $\sin(n\pi(t-\tau)/T)$ with $n=1,2,\cdots$. Since $\frac{\partial}{\partial{t}}f_{k\neq{\tilde{k}}}(\mathbf{x}_{1\textrm{-}},t)$ satisfies the same conditions as $y(t)$, the time dependent part of $\frac{\partial}{\partial{t}}f_{k\neq{\tilde{k}}}(\mathbf{x}_{1\textrm{-}},t)$ can be expanded by those basis functions, i.e.,
\begin{equation}
\setlength\abovedisplayskip{0pt}
\setlength\belowdisplayskip{0pt}
f_{k\neq{\tilde{k}}}(\mathbf{x}_{1\textrm{-}},t)=\sum_nD_{k\neq{\tilde{k}},n}(\mathbf{x}_{1\textrm{-}})\cos(n\pi(t-\tau)/T)
\end{equation} 
where $D_{k\neq{\tilde{k}},n}(\mathbf{x}_{1\textrm{-}})$ represent the expansion coefficients, thus rendering 
\begin{equation}
\setlength\abovedisplayskip{0pt}
\setlength\belowdisplayskip{0pt}
f_{k\neq{\tilde{k}}}(\mathbf{x}_{1\textrm{-}},t+2T)=f_{k\neq{\tilde{k}}}(\mathbf{x}_{1\textrm{-}},t).
\end{equation} 
Equation (45) clearly shows the $2T$ period of the functions $f_{k\neq{\tilde{k}}}(\mathbf{x}_{1\textrm{-}},t)$.

\section{Discussions}

\subsection{Consistent interpretation of the reduced density matrix}

The reduced density matrix is defined with respect to the preferred-basis of a particular sub-system. For convenience, denote such matrix by $\mathbf{M}$ with elements $\mathbf{M}_{k,k'}$. Using sub-system one as an example and keeping the definition $\mid\phi_{1,k}(\mathbf{x}_{1\textrm{-}},\tau)>=<\psi_{1,k}(\mathbf{x}_1)\mid\psi(\mathbf{x},\tau)>$ in mind, the elements at time $t=\tau$ are 
\begin{equation}
\setlength\abovedisplayskip{3pt}
\setlength\belowdisplayskip{0pt}
\mathbf{M}_{k,k'}=<\phi_{1,k}(\mathbf{x}_{1\textrm{-}})\mid\phi_{1,k'}(\mathbf{x}_{1\textrm{-}})>.
\end{equation}
According to the decoherence program [3-10], appropriate interaction between the sub-system and the environment can make the off-diagonal elements $\mathbf{M}_{k,k'\neq{k}}$ near zero. That program also suggested that the wave function will not start collapsing unless the off-diagonal elements are small, which is referred to as the decoherence condition. In other words, the off-diagonal elements affect the chance to start a collapse process.  

The values of the diagonal elements $\mathbf{M}_{k,k}$ represent the probability for the sub-system to collapse to the basis function $\mid\psi_{1,k}(\mathbf{x}_1)>$. Those probabilities can be any value between zero and one. To maintain consistence, the off-diagonal elements should also represent some sort of probability ranging from zero to one. Keeping the decoherence condition in mind, it is reasonable to assert that the off-diagonal elements represent the probability to transition from Schr\"odinger equation to the collapse equation.  

The probabilistic meaning of the off-diagonal elements requires the reduced density matrix to be checked at discrete time points. Otherwise the probability to start a collapse process can only be zero or one. Given the matrix and the time points, the cumulative probability to collapse the wave function in any interval can be calculated. In case the matrix repeats its value after some time, the cumulative probability to get collapse processes in an earlier interval should equal that in a later interval as long as the two intervals are equal. To ensure this result irrespective of the value of the matrix, the lapses between any two adjacent time points have to be constant.

\subsection{Alternating evolutions}
 
Equations (12) and (23) together describes the only valid solution for the collapse equation (5). The solutions of (12) or the periodic functions (39), which not only explain the origin of some preferred-basis but also enable the consistent interpretation of the reduced density matrix, are derived under the condition that the collapse equation (5) smoothly transitions to Schr\"odinger equation (1). The significance of (39) and no need of new principle to derive (39) demonstrate the elegance of the presented dynamics of collapse. 

The periodic function defines two sets of intervals. One set of intervals include the half cycle $[\tau, \tau+T]$ and those that are integer number of cycles away from it. The rest intervals are in the other set. The collapse equation may guide the evolution of the wave function only during the set of intervals that contain $[\tau, \tau+T]$. At the beginning of each interval, the probability to start a collapse process is defined by the off-diagonal elements in the reduced density matrix. If there is no qualified collapse, the wave function will continue to follow Schr\"odinger equation. Otherwise, the evolution scheme will switch to the collapse equation at the beginning of the interval and smoothly transition back to Schr\"odinger equation after time $T$.

The universe consists of almost infinite number of sub-systems. Therefore, there may always exist large number of sub-systems whose reduced density matrices demand non-zero probabilities of collapse. Consequently, the evolution may always switch to the collapse equation at the beginning of the qualified intervals. To see this by an example, assume there are 1000 sub-systems that all have $1\%$ chance to collapse. The probability not to start a collapse process will be $0.99^{1000}$ or 0.000043. That means it is almost sure that Schr\"odinger equation and collapse equation take turn to guide the wave function of the whole universe for half cycle each turn. 

\subsection{The simplification of the periodic functions}
All the periodic functions satisfy the same conditions (40-43). Considering the arbitrariness of sub-system 1 and the system, those functions have to be identical and only depend on time. That's to say, $\lambda_{k\neq{\tilde{k}}}^{-1}(\mathbf{x}_{1\textrm{-}},t)$ and $f_{k\neq{\tilde{k}}}(\mathbf{x}_{1\textrm{-}},t)$ should be $\lambda^{-1}(t)$ and $f(t)$ respectively. 

Assume that $\lambda_{\tilde{k}}^{-1}(\mathbf{x}_{1\textrm{-}},t)$ does not depend on $\mathbf{x}_{1\textrm{-}}$. Left-multiplying $<\psi(\mathbf{x},t)\mid$ to both sides of (5) yields
{\setlength\arraycolsep{0pt}
\begin{eqnarray}
\fl <\psi(\mathbf{x},t)\mid{i}\hbar\frac{\partial}{\partial{t}}\mid{\psi}(\mathbf{x},t)> =i\hbar\sum_{k\neq{\tilde{k}}}\lambda^{-1}(t)||<\psi_{1,k}(\mathbf{x}_1)\mid\psi(\mathbf{x},t)>||^2\nonumber\\
\fl +i\hbar\lambda_{\tilde{k}}^{-1}(t)||<\psi_{1,\tilde{k}}(\mathbf{x}_1)\mid\psi(\mathbf{x},t)>||^2
+<\psi(\mathbf{x},t)\mid{H}(\mathbf{x},t)\mid\psi(\mathbf{x},t)>. 
\end{eqnarray}
Making use of $\sum_{k\neq{\tilde{k}}}||<\psi_{1,k}(\mathbf{x}_1)\mid\psi(\mathbf{x},t)>||^2=1-||<\psi_{1,\tilde{k}}(\mathbf{x}_1)\mid\psi(\mathbf{x},t)>||^2$, (47) becomes
{\setlength\arraycolsep{0pt}
\begin{eqnarray}
\fl <\psi(\mathbf{x},t)\mid{i}\hbar\frac{\partial}{\partial{t}}\mid{\psi}(\mathbf{x},t)> =i\hbar\lambda^{-1}(t)(1-||<\psi_{1,\tilde{k}}(\mathbf{x}_1)\mid\psi(\mathbf{x},t)>||^2)\nonumber\\
\fl +i\hbar\lambda_{\tilde{k}}^{-1}(t)||<\psi_{1,\tilde{k}}(\mathbf{x}_1)\mid\psi(\mathbf{x},t)>||^2
+<\psi(\mathbf{x},t)\mid{H}(\mathbf{x},t)\mid\psi(\mathbf{x},t)>. 
\end{eqnarray}
To conserve energy during collapse, it follows from (48) that
\begin{equation}
\setlength\abovedisplayskip{0pt}
\setlength\belowdisplayskip{0pt}
\lambda_{\tilde{k}}^{-1}(t)=\lambda^{-1}(t)(1-||<\psi_{1,\tilde{k}}(\mathbf{x}_1)\mid\psi(\mathbf{x},t)>||^{-2}).
\end{equation} 
Therefore the collapse equation (5) can be re-written as
{\setlength\arraycolsep{0pt}
\begin{eqnarray}
\fl i\hbar\frac{\partial}{\partial{t}}\mid{\psi}(\mathbf{x},t)> =i\hbar\lambda^{-1}(t)(1-||<\psi_{1,\tilde{k}}(\mathbf{x}_1)\mid\psi(\mathbf{x},t)>||^{-2}\mid\psi_{1,\tilde{k}}(\mathbf{x}_1)><\psi_{1,\tilde{k}}(\mathbf{x}_1,t)\mid)\nonumber\\
\mid\psi(\mathbf{x},t)>+H(\mathbf{x},t)\mid\psi(\mathbf{x},t)> 
\end{eqnarray}
where 
\begin{equation}
\setlength\abovedisplayskip{0pt}
\setlength\belowdisplayskip{0pt}
\lambda^{-1}(t)=\frac{f(t)-1}{f(t)}
\end{equation} 
with
\begin{equation}
\setlength\abovedisplayskip{0pt}
\setlength\belowdisplayskip{0pt}
f(t)=\sum_n{D_n}\cos(n\pi(t-\tau)/T).
\end{equation} 
Equations (51) and (52) are simplified (39) and (44) respectively without the dependence on $\mathbf{x}_1$ and $k$.

\subsection{The collapse equation for multiple sub-systems}

Equation (50) describes the collapse of one sub-system. It needs to be generalized to handle multiple collapses. For conciseness without loss of generality, it suffices to consider two collapses, one for sub-system 1 to collapse to $\mid\psi_{1,\tilde{k}}(\mathbf{x}_1)>$ and the other for sub-system 2 to collapse to $\mid\psi_{2,\tilde{m}}(\mathbf{x}_2)>$. Using the same method as in deriving (50), the following dynamic equation is obtained
{\setlength\arraycolsep{0pt}
\begin{eqnarray}
\fl i\hbar\frac{\partial}{\partial{t}}\mid{\psi}(\mathbf{x},t)> =i\hbar\lambda^{-1}(t)(2-||<\psi_{1,\tilde{k}}(\mathbf{x}_1)\mid\psi(\mathbf{x},t)>||^{-2}\mid\psi_{1,\tilde{k}}(\mathbf{x}_1)><\psi_{1,\tilde{k}}(\mathbf{x}_1)\mid\nonumber\\
 -||<\psi_{2,\tilde{m}}(\mathbf{x}_2)\mid\psi(\mathbf{x},t)>||^{-2}\mid\psi_{2,\tilde{m}}(\mathbf{x}_2)><\psi_{2,\tilde{m}}(\mathbf{x}_2)\mid)\mid\psi(\mathbf{x},t)>\nonumber\\
+H(\mathbf{x},t)\mid\psi(\mathbf{x},t)>. 
\end{eqnarray}

\section{Conclusions}
  
This paper first presented the collapse equation that has to contain a cyclic function to guarantee smooth transition between collapse process and Schr\"odinger evolution. The derived collapse equation adds some constraints to preferred-basis and supports the consistent interpretation of the reduced density matrix, i.e., diagonal elements represent probabilities to collapse to basis functions while off-diagonal elements represent the probability to start a collapse process. Its cyclic function defines half cycle intervals for the evolution schemes. Due to the enormous number of sub-systems in the universe, it is almost certain that either evolution scheme is only valid for half cycle before it transitions to the other type. This paper also presented the unique set of determination equations of preferred-basis that yield solutions able to explain many quantum measurement results.

\end{document}